# Mirror-time diffusion discount model of options pricing

Pavel Levin[1]

*St. John's University, Physics Department, 8000 Utopia Pkwy, Queens, NY 11439 USA*

**Abstract**

The proposed model modifies option pricing formulas for the basic case of log-normal probability distribution providing correspondence to formulated criteria of efficiency and completeness. The model is self-calibrating by historic volatility data; it maintains the constant expected value at maturity of the hedged instantaneously self-financing portfolio. The payoff variance dependent on random stock price at maturity obtained under an equivalent martingale measure is taken as a condition for introduced "mirror-time" derivative diffusion discount process. Introduced $\xi$-returns, correspondent to the found general solution of backward drift-diffusion equation and normalized by theoretical diffusion coefficient, don't contain so-called "long tails" and unbiased for considered 2004-2007 S&P100 index data. The model theoretically yields skews correspondent to practical term structure for interest rate derivatives. The method allows increasing the number of asset price probability distribution parameters.

---

[1] Tel./fax: +1 718 645-9366. Email: *levinpavel@yahoo.com*

# 1. Introduction

In the standard Black-Scholes-Merton option pricing model [1, 2], the delta-hedged portfolio growth determines the diffusive partial differential equation in the underlying price-time coordinates (BSM PDE). The famous BSM formula can be derived as the PDE particular solution with a terminal condition represented by a payoff function, or as a discounted expected derivative value at maturity obtained under martingale measure **Q**, equivalent, according to Girsanov [3], to the real-world log-normal probability measure **P** (see also Wilmott e.a. [4]). The BSM PDE asserts upholding constant hedged portfolio value for given payoff function, at that the condition at maturity is denoted by a real variable. Using the real variable in condition expression is inconsistent with a formulation of underlying stochastic process with a random variable at arbitrary time. On the other hand, employing random-variable condition at maturity would contradict to requirement of upholding the constant portfolio value at arbitrary time. The widely-used exponential discount from expected payoff implicitly assumes that a derivative value forwarded to maturity is a martingale under both measures **P** and **Q**, which is true for an underlying security (or forward) but not for an option with asymmetric payoff function. As can be shown, discounting the derivative value back to the current time under the real-world measure **P** rather than under appropriate log-normal-world martingale measure **Q** causes the bias between implied and historical volatilities; that undermines the model efficiency. In our opinion, for corresponding the efficient market hypothesis, the BSM model requires modification applying Girsanov's equivalent martingale measure to derivation of both expected derivative value at maturity and its value discounted to the current time.

The delta-hedging is practically feasible for local volatility models of Dupire [5], Rubinstein [6], describing the diffusion process with variable coefficients, which can be recovered from conditional probability density ("volatility smile" level, slope, and curvature). However, theoretically, according to Ait-Sahalia e.a. [7], the differences between the stock and option implied risk-neutral densities within the framework of BSM diffusion with exponential discount ultimately would lead to the pricing inefficiency.

The determination of diffusion coefficients is complicated by well-known fact that assumed log-normality is violated in stock return distribution time-series. Besides an empirical phenomenon called "volatility smile" in option markets, the leptokurtic feature takes place. The return distribution of assets may have a higher peak and asymmetric tails, heavier than those of the normal distribution; this led many authors to consider jump-diffusion models with Levy flights (first proposed by Merton [8]). For example, Kou [9] assumes a double-exponential conditional distribution for the jump size; such many-parameter model is sufficient for description of the volatility smile parameters. The models with increased number of parameters price options across strikes and maturities more accurately; however, the issue of parameter stability arises. The mentioned long tails can be eliminated by using normalized distributions introduced below.

Analogously to equity derivatives, the fixed income options are priced by Black [10] formula as an exponentially discounted to the current time expected payoff value at maturity for the case of log-normally distributed forward price. Within the framework of Heath-Jarrow-Morton [11] term structure of interest rates expressed as functions of their volatilities, the bond and its derivative prices at arbitrary time are determined by exponential discount with integral-average rate for the period to maturity. Brace-Gatarek-Musiela [12] Libor forward rate structure model describes the dynamics of a family of forward rates under a common measure. But, unlike stocks, the interest rate futures are derivatives. The existing differences between implied volatilities of two derivative types - the interest rate futures and options, according to de Jong e.a. [13], should theoretically lead to the possibility of arbitrage. According to Gupta and Subrahmanyam [14], for improving a pricing accuracy of interest rate options, there is a need for introducing a second stochastic factor, mean-reversion coefficient determining the term structure evolution through time. For consistent pricing and hedging, a further increasing of a number of parameters is suggested, however, at expense of model stability and extensive computation resources.

The general stochastic volatility models of Heston [15], Hull and White [16] introduce an additional stochastic process for underlying security's volatility, governed by its price level. It allows introducing necessary corrections to exponentially discounted expected price value, which is dependent on volatility. However, improving the pricing accuracy is achieved at expense of the model completeness (ability to hedge options with the underlying asset) as an additional degree of freedom and the market price of volatility risk was introduced. For providing pricing efficiency, these many-parameter models require frequent calibration against historical underlying security data with consequent fitting to an actual smile, minimizing residual errors.

Carr and Madan [17] showed that the absence of call spread, butterfly spread and calendar spread arbitrages for the Markovian assumption is sufficient for exclusion of all static arbitrages from a set of option price quotes across strikes and maturities on a single underlying. No-arbitrage conditions imply the risk-neutral probability measure conservation law for a *stock price variance at maturity* and a possibility of hedging during *time-to-maturity,* using static position in a set of available options for any nearer maturity.

The present paper proposes an efficient option pricing model based on a BSM modification maintaining the constant hedged portfolio expected value at maturity. We take into account supposed stock price and payoff variance at maturity obtained under an equivalent martingale measure, making it conditional for introduced derivative diffusion discount process during time-to-maturity considered also under an equivalent martingale measure.

## 2. Generalized differential equation of options pricing

*2.1. Complete market and self-financing portfolio adjusted to maturity*

Let's consider geometric Brownian motion for an asset price $S_t$ in $\boldsymbol{\Omega}$-space with filtration $\mathfrak{I}_t^P$ under probability measure **P** with stochastic differential consisting of the drift term and the Wiener process, $W_t$, in the relative time $t$ within the interval corresponding to one in absolute time from instant $\tau$ to an asset derivative maturity $T$:

$$dS_t = \mu(t_0, 0)S_t dt + \sigma(t_0, 0)S_t dW_t, \quad t \in [t_0, 0], \quad t_0 = \tau - T < 0. \tag{1}$$

The considered case is provided by time-invariant diffusion with an averaged volatility $\sigma$ and a drift rate $\mu = r - q$, where, $r$ is the averaged risk-free interest rate; $q$ - continuous dividend rate for the asset. According to the efficient market hypothesis, at any relative time $t$, given drift and volatility parameters, the current asset price $S_t$ incorporates instantaneously any information concerning the market future evolution up to the derivative maturity.

The market is *arbitrage-free*, i.e. at any relative time $t < 0$ the asset price can be adjusted to maturity ($t = 0$) with price deflator $\exp(\mu t)$:

$$S_t^T = S_t \exp(-\mu t), \tag{2}$$

which is a martingale, with expected value at maturity determined by initial condition $S_{t=t_0} = S_0$:

$$S_0^T \equiv S_0 \exp(-\mu t_0) = E_{t_0,S_0}^P \langle S_{t=0} | \mathfrak{I}_t^P \rangle \equiv E_{t_0,S_0}^P [S_{t=0}]. \tag{3}$$

We define the market as *complete* at relative time *t* if, for a contingent claim (option) on the asset (stock) $V_t = V(S_t)$, there exists a self-financing trading strategy such that the expected value at maturity of a portfolio consisting of a long option position and short Δ stocks remains unchanged:

$$E_{t,S}^P [\Pi_{t=0}] \equiv E_{t,S}^P [V(S_{t=0}) - \Delta(S_{t=0})S_{t=0}] = \Pi(t_0)\exp(-rt_0) \equiv V^T(S_0) - \Delta(S_0)S_0^T, \tag{4}$$

$$V^T(S_t) = V(S_t)\exp(-rt). \tag{5}$$

The above definition of complete market imposes additional constraint on the *instantaneously self-financing* (growing with risk-free interest rate *r*) portfolio, $\Pi(t) \equiv V(S_t) - \Delta(S_t)S_t$. Since the product $\Delta(S_t)S_t^T$ is not a martingale for non-symmetric $\Delta(S_t) = \partial V_t / \partial S_t$ functions, and an adjusted to maturity option value $V^T(S_t)$ is never a martingale for real payoff functions. Therefore, the said definition is not identical and is complementary to the usual definition of market completeness at instant *t*.

*2.2. Delta-hedged portfolio and options pricing PDE adjusted to maturity*

The hedging strategy corresponding to the above definition of completeness considers self-financing portfolio $\Pi$, analogous to BSM, in which position on the derivative is delta-hedged against the risk of the random asset price depreciation. However, as far as $V^T(S_t)$ in (4) is not a martingale, the option expected value at maturity $E^P_{t,S}[V(S_{t=0})]$ cannot be just discounted with deflator $\exp(rt)$ like in known models; one can operate only with current prices $V_t$, $S_t$.

Let's consider the derivative value discounted from maturity ($t = 0$) to arbitrary moment $t' \in [t, 0]$. For the "dummy" random price $S'_{t'}$, unchanged expected portfolio value at maturity (4) can be taken as a terminal condition determined by payoff $\Phi(S_{t=0}, K)$ at strike $K$ and maturity $t' = 0$:

$$d\Pi \equiv d'V(S'_{t'}) - \frac{\partial V_t}{\partial S'_{t'}} dS'_{t'} = r\Pi(t')dt', \qquad (6)$$

$$S'_{t'=t} = S_t, \quad S'_{t'=0} = S_{t=0}, \quad V(S'_{t'=0}) = \Phi(S'_{t'=0}, K), \qquad (7)$$

$$E^P_{t',S}[\Phi(S'_{t'=0}, K) - \left(\frac{\partial \Phi}{\partial S'_{t'}}\right)_{t'=0} S'_{t'=0}] = V^T(S_0) - \Delta(S_0) S_0^T. \qquad (4')$$

If $S'_{t'}$ process is also determined as a geometric Brownian motion (1), conditions (4) and (4') become incompatible for portfolio (6) unless the "dummy" discounted option price at arbitrary time $t'$ is specified as conditional on random asset value at maturity evaluated at relative time $t$:

$$V_{t'} = V\langle S'_{t'} | S_{t=0} \rangle. \qquad (8)$$

The introduced "dummy" inversed discount process $S'_{t'}$ is directed from maturity to arbitrary moment $t' < 0$ and determined not only by drift rate and the Wiener process, but also by supposed asset price distribution at maturity, $S_{t=0}$. So-called "risk-neutral" discount of $V_{t=0}$ value may be obtained as a result of continuous hedging against the "dummy" discount process $S'_{t'}$ such that the hedged portfolio upholds its terminal random value at maturity. Due to information available at relative time $t$, the hedging dynamics can be established in so-called "mirror" *time-to-maturity* $|t'|$, as a function of a stochastic process $S'_{t'}$ mirrored to the underlying process $S_t$ and specified within the probability space ($\mathbf{\Omega}$, $\mathfrak{I}^P_{|t'|}$, $\mathbf{P}$). The stochastic differential of the option pricing function $V_{t'}$ can be expressed according to Ito's lemma:

$$dS'_{t'} = \mu(t', 0)S'_{t'}dt' + \sigma(t', 0)S'_{t'}dW_{|t'|}, \quad |t'| \in [0, |t|]; \tag{9}$$

$$d'V_{t'} = \left(-\frac{\partial V_{t'}}{\partial |t'|} + \mu S'_{t'}\frac{\partial V_{t'}}{\partial S'_{t'}} + \frac{\sigma^2 S'^2_{t'}}{2}\frac{\partial^2 V_{t'}}{\partial (S'_{t'})^2}\right)dt' + \sigma S'_{t'}\frac{\partial V_{t'}}{\partial S'_{t'}}dW_{|t'|}. \tag{10}$$

While the deterministic part of discount process $S'_{t'}$ is expressed in direct time $t'$ and undistinguishable from one of the underlying process $S_t$, the stochastic parts of processes (9), (10) are given in the "mirror" time $|t'|$. Putting (6), (9), (10) together, one can eliminate the relative time differentials, $dt'$, and the Wiener differentials, $dW_{|t'|}$. Using the option price adjusted to maturity (5), the strategy of hedging in "mirror" time $|t'|$ can be expressed by derivative discount PDE:

$$\frac{\partial V^T_{t'}}{\partial |t'|} = \frac{\sigma^2 S'^2_{t'}}{2}\frac{\partial^2 V^T_{t'}}{\partial S'^2_{t'}} + rS'_{t'}\frac{\partial V^T_{t'}}{\partial S'_{t'}}. \tag{11}$$

PDE (11) expresses drift-diffusion in the moving logarithmic coordinates $S'_{t'}$ with the system drift rate $r$, analogous to well-known BSM PDE in inversed time (see Wilmott [4]). However, in the standard BSM PDE, despite the time reversion to $-t$, the stochastic parts of both processes - for stock $S_{-t}$ and for derivative $V_{-t}$ - remain being co-directed in time, from *t* to maturity. By simple time reversing in (1), the resulting direction of the diffusion expressed by Wiener process cannot be changed ($-dW_{-t} = dW_t$):

$$dS_{-t} = -\mu(t_0, 0)S_{-t}dt - \sigma(t_0, 0)S_{-t}dW_{-t}, \quad -t \in [0, -t_0].  \quad (1')$$

The unchanged direction of the stochastic part of process (1') is just opposite to the drift-diffusion direction stated by BSM PDE in inversed time and supposed in BSM solution. Moreover, the sign ought to be changed not only for the time-derivative BSM PDE term, but also for the gradient term containing the system velocity (i.e. the drift rate $\mu$ should change sign, too - see Carslaw and Jaeger [18]); that would lead to irrelevant results. In contrast to the simple time-inversion for underlying process (1), introducing the separate "dummy" process in the "mirror" time-to-maturity $|t'|$ ($t' < 0$; $dW_{|t'|} = -dW_{t'}$), really changes direction of the stochastic part of $V_{t'}$-diffusion, as expressed in (9), (10).

*2.3. The solution of derivative discount PDE*

PDE (11) governs geometric Brownian motion, which marginal density function corresponds to log-normal distribution. It can be solved after well-known logarithmic coordinate transformation, related here to the asset price adjusted to maturity (2):

$$\zeta_{t'}^{Q} = \ln(S'_{t'}/S_{t'}^{T}) + (\mu - \sigma^2/2)|t'|, \quad (12)$$

Transformation (12) leads to equivalent standard diffusion equation reducing stochastic process $V_{t'}^{T}$, which is *not* a martingale under **P** and, therefore, is *not* identical to expected payoff, $E_{t',S'}^{P}[V(S'_{t'=0})]$, from maturity back to time $t'$ under equivalent measure **Q** (filtration $\Im^{Q}$, with diffusion coefficient *D*):

$$\frac{\partial V_{t'}^T}{\partial |Dt'|} = \frac{\partial^2 V_{t'}^T}{\partial (\zeta_{t'}^Q)^2} . \tag{13}$$

An equivalent PDE (13) describes a normal distribution in $(t', \zeta_{t'}^Q)$-space, which corresponds to the log-normal probability distribution in $(t', S_{t'}')$ and $(t, S_t)$-spaces. The stock price variance at maturity is determined by its probability density, a function of random variable $S_{t=0}$ with its currently expected value $S_t^T$ as a parameter:

$$f_{t,S}^Q(S_{t=0}) = \frac{1}{S_{t=0}\sqrt{2\sigma^2|t|}} \exp\left\{-\frac{[\log(S_{t=0}/S_t^T) + \sigma^2|t|/2]^2}{2\sigma^2|t|}\right\} . \tag{14}$$

The stochastic process $V_{t'}^T$ is a martingale under equivalent probability measure **Q** with correspondent expected value:

$$V_{t'}^T \equiv E_{t',S'}^Q[V(S_{t=0})] = \int_0^\infty \Phi(S_{t=0}, K) f_{t,S}^Q(S_{t=0}) dS_{t=0} . \tag{15}$$

Substitution of log-normal probability density (14) into integral (15) yields famous BSM formula, which represents an expected option value at maturity and initial condition for inversed "mirrored" discount processes (9), (10) and correspondent PDE (11). The expected option value at maturity $E_{t,S}^Q[V(S_{t=0})]$ can be related to the currently expected underlying value $S_t^T$ by means of an existing equivalent martingale measure **Q**; the same is true for the option value $V_{t'}$ discounted from maturity to arbitrary time $t'$.

The initial condition of PDE (11) and equivalent PDE (13) represented by integral (15) includes payoff function $\Phi$ of random variable, $S_{t=0}$, and not identical to payoff function of the real variable, which is commonly accepted for the option pricing PDE in inversed time [4]. Generally saying, $d'V$ in (10) could be considered as *a total* *differential only if the stock price variance at maturity is considered. The possibility of hedging taking into account the "mirror-time" derivative discount during given time-to-maturity period as well as the stock price variance at maturity adds up to an option value.*

Analogously to the general solution of equivalent PDE (13) (see Carslaw and Jaeger [18]), the one of correspondent PDE (11) can be found according to Wilmott e.a. [4] as an integral containing initial condition in the form of a function of "dummy" variable, $V^T(S'^T)$:

$$V^T(S'_{t'}) = \frac{1}{\sigma\sqrt{2\pi|t'|}} \int_0^\infty \exp\left(-\frac{\left(\ln(S'^T/S'^T_{t'}) + \sigma^2|t'|/2\right)^2}{2\sigma^2|t'|}\right) V^T(S'^T) \frac{dS'^T}{S'^T}. \qquad (16)$$

The particular solution for payoff function (15) for the case of interest, $t = |t'| = t_0$, $S_{t=t_0} = S_0$, $S_0^T \equiv S_0\exp(-\mu t_0)$, takes the form:

$$V^T(S_0) = \frac{1}{\sigma\sqrt{2\pi|t_0|}} \int_0^\infty \exp\left(-\frac{\left(\ln(S'^T/S_0^T) + \sigma^2|t_0|/2\right)^2}{2\sigma^2|t_0|}\right) E^Q_{t_0,S_0}[V(S'^T)] \frac{dS'^T}{S'^T}, \qquad (17)$$

*2.4. The equivalent martingale measure and correspondent coordinate transformation*

The found general solution (16) contains an internal integral (15), which can be reduced to commonly used BSM solution, and the kernel function corresponding to the fundamental solution of drift-diffusion PDE (11). An equivalent PDE (13) is written in $\zeta^Q$-coordinates defined by transformation (12) explicitly containing time; that's not always convenient for econometric studies. The PDE (13) in an equivalent time-to-maturity $|Dt|$ can be obtained also by coordinate transformation not containing explicitly time accordingly to method proposed by Levin [19]:

$$\xi = (S/S_0)^{1-2\mu/\sigma^2}, \tag{18}$$

$$D = \frac{\sigma^2}{2}\left(1-\frac{2\mu}{\sigma^2}\right)^2 \left(\frac{S}{S_0}\right)^{2\left(1-\frac{2\mu}{\sigma^2}\right)} \tag{19}$$

According to (18), (19), *the normalized $\xi$-returns* corresponding to an equivalent martingale measure **Q**, can be expressed as $(\xi(\tau_{i+1})-\xi(\tau_i))/\sqrt{D(\tau_{i+1}-\tau_i)}$, with diffusion coefficient $D$ incorporating volatility $\sigma$ and drift rate $\mu$. Comparatively to widely used normalized log-return expression, $(\zeta(\tau_{i+1})-\zeta(\tau_i))/(\sigma\sqrt{\tau_{i+1}-\tau_i})$, $\zeta = \ln(S/S_0)$, the normalized $\xi$-return, is more convenient for econometric studies at sufficient drift rates.

## 3. The model verification and discussion

*3.1. Design of study*

For practical European call equity and future options valuation, using particular solution (17) for the case of normal distribution, one can obtain formula in terms of time to maturity $T$ and adjusted forward price $S_0^T$ (in this section we consider $\tau = 0$):

$$V^C = \frac{e^{-rT}}{\sigma\sqrt{2\pi T}} \int_K^\infty \exp\left(-\frac{\left(\ln(S'^T/S_0^T)+\sigma^2 T/2\right)^2}{2\sigma^2 T}\right)[S'^T N(d_1) - KN(d_2)]\frac{dS'^T}{S'^T},$$

$$d_1 = \frac{\ln(S'^T/K)+\sigma^2 T/2}{\sigma\sqrt{T}}, \tag{20}$$

$$d_2 = d_1 - \sigma\sqrt{T},$$

where, $N(.)$ is the standard cumulative normal distribution function, $S'^T$ is a dummy forward price (variable of integration).

After a computation of the call price $V^C$ according to the proposed model for given historical volatility $\sigma$ for correspondent period, we express it in the terms of Black-Scholes implied volatility, *SigI_Model*, and compare with the published data on implied volatility, *SigI_Data*, for given maturity, *T*, and relative moneyness,

$k = (S_0^T - K)/S_0^T$ .

The return probability distribution was found for $\xi$-returns (18), *Ksi/DIV/Sqrt(2t)*, normalized with current theoretical diffusion coefficient, *DIV* (19), with substituted model implied volatility data *SigI_Model*, corresponding to theoretical option price (20) based on 90-day historical volatility data. For comparison, the return probability distribution was expressed for $\xi$-returns normalized with current 30-day historical volatility, *Ksi/D30/Sqrt(2t)*, as well as for normalized log-returns, *Zta/Sig30/Sqrt(2t)*.

*3.2. S&P 100 data analysis based on generalized solution*

Daily return data and modelling results for S&P 100 option prices with 30-day maturity for month/year period 4/2004 – 4/2007 are given in Fig. 1. Calculated return distributions of daily $\xi$-returns normalized with current 30-day historical volatility and normalized log-returns for considered annual drift rate range 1…5.25% are negligibly distinguishable (Fig. 1, a); that's because the average daily drift 0.0134% is much less than average daily S&P 100 index volatility corresponding to average yearly volatility *Sig30*=10.3%. The difference between distributions for $\xi$-returns and log-returns could be seen only for much larger periods (quarters and years), which are off the limits of the present paper.

The density distributions for both types of normalized returns sufficiently deviate from the normal distribution (thin solid curve). The asymmetric leptokurtic features takes place: the return distribution is skewed to the right and has two heavier tails than those of the normal distribution. It's interesting to note that the long tails largely correspond to normal distribution characterized by average Black-Scholes implied volatility *SigI*=13.2% (see Fig. 1, a, dashed curve).

The probability distribution for S&P 100 $\xi$ -returns normalized with the theoretical diffusion coefficient, *Ksi/DIV/Sqrt(2t)*, also shows a peak skewed to the right; however, the tails largely correspond to ones of normal distribution (Fig. 1, b). While deviations from normal probability distribution (skewed peak) still take place for returns normalized by modeled diffusion coefficient and could be addressed by introducing additional distribution parameters, such "long-tail-less" distribution implies no need, for example, in Levy flights characterizing jump-diffusion process.

BSM results based on 30-day and 90-day historical volatility *SigH30, SigH90* and the present diffusion-discount model results for European call at the money based on *SigH90* and expressed in terms of BS implied volatility *SigI_Model* were compared with VXO implied volatility index *SigI_Data* (Fig. 2). In distinction to historical volatility comparison to BS implied volatility, the results of diffusion-discount model for normal distribution (determined by formula (20) with $K = S_0^T$ ) are practically unbiased relatively implied volatility data (bias 1.85% comparatively to 21.1% for historical volatility). For given S&P 100 index 4-year data, the developed model appears to be self-calibrating with no need in setting up additional jump-diffusion and stochastic volatility parameters. The model accuracy, including volatility skew ("smile") calculations for equity options, may be improved by increasing the number of the probability distribution parameters in (14) and introducing variable diffusion coefficients in discount option pricing PDE (11), (13).

*3.3. Pricing analysis for interest income derivatives*

Prices of fixed income options (caps and swaptions) contain information about interest rate volatilities and correlations, which can be inverted in the framework of Heath- Jarrow-Morton [11] and Brace-Gatarek-Musiela [12] models to the option-implied interest rate volatility term structure. The empirical analysis based on weekly US 1995-1999 data showed that the option-implied Black volatility is sufficiently higher than correspondent zero-coupon-bond-based forward interest rate volatility. Such a bias for different fixed income derivatives on the same underlying could theoretically lead to the possibility of arbitrage, see Jong e.a. [13]. That's why proposed model result examination for interest income derivatives is of special interest.

Using formula (20) for different forward interest rate-based volatilities *SigI_Forw*, we calculated forward prices in terms of Black implied volatility for caps *SigI_Model*; results are given in Fig. 3. Calculations results for different maturities (Fig. 3, a) explain the difference between interest-rate-based and option-implied term structure. The difference *(SigI_Model - SigI_Forw)* and skewness of calculated *SigI_Model* for different relative moneynesses (Fig. 3, b) correspond to the pricing error for caps reported by Gupta and Subrahmanyam [14].

The proposed model modifies basic log-normal-distribution pricing formulas comparatively to the Black model, explaining the volatility skew and sufficiently improving efficiency for different maturities and moneynesses of interest income

4options. Achieving efficiency of one-parameter derivatives pricing model is a prerequisite of its stability at increasing the number of distribution parameters.

**Conclusions**

The proposed one-parameter option pricing model modifies formulas for the basic case of log-normal probability distribution providing correspondence to formulated criteria of efficiency and completeness. The model is self-calibrating by historic volatility data; it maintains the constant expected value at maturity of the hedged instantaneously self-financing portfolio. For this instance, it takes into account the random stock price and payoff variance at maturity obtained under an equivalent martingale measure as conditional for introduced "mirror-time" derivative diffusion discount process. The "risk-neutral" discount of derivative value may be obtained as a result of continuous hedging against the inversed "dummy" process directed back from maturity and specified within the same probability space as an underlying process. The possibility of hedging taking into account the derivative diffusion discount during time-to-maturity period as well as the stock price variance at maturity adds up to an option value. The found general solution of correspondent backward drift-diffusion PDE contains an internal integral, which can be reduced to commonly used BSM solution, and the kernel function corresponding to its fundamental solution. Introduced normalized $\xi$-returns are not dependent explicitly on time and thus convenient for using in econometric studies at sufficient drift rates. While the probability distribution for $\xi$-returns normalized with theoretical diffusion coefficient shows some deviations from normal probability distribution (skewed peak), which can be addressed by introduction of additional distribution parameters, the tails largely correspond to ones of normal distribution; thus, there is no need in consideration of jump-diffusion. For given S&P 100 index 4-year data, the developed one-parameter model appears to be efficient with bias only 1.85% comparatively to 21.1% for BSM model; therefore, there is no need in introducing stochastic volatility parameters and the model frequent calibration. The proposed model results allow matching the forward interest rate volatility structure with that of implied volatility of interest rate derivatives (such as caps). The model theoretically explains the implied volatility skew, and can be used for practical pricing of options with different strikes and maturities. Achieving efficiency of one-parameter option pricing model is a prerequisite of its stability at increasing the number of distribution parameters.

**Acknowledgements.**

The author thanks M. Avellaneda and A. Lipton for useful discussions on the initial stage of this work.

**Figures**

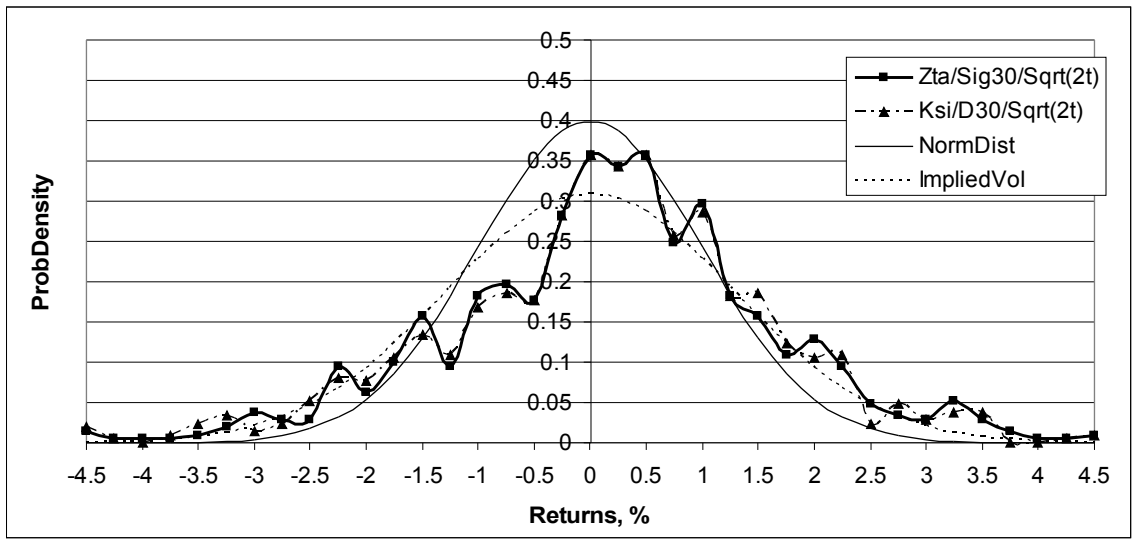

a

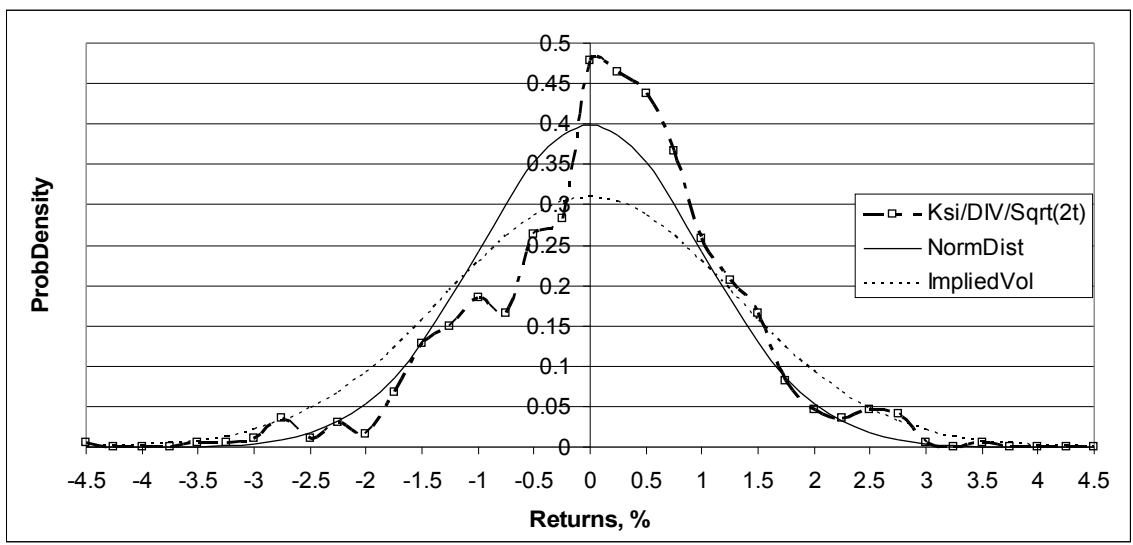

b

Fig.1. S&P 100 options pricing with 30-day maturity for period 4/2004 –4/2007: the index return distribution - log-returns and $\xi$ -returns normalized with 30-day historical volatility (a); $\xi$ -returns normalized with theoretical diffusion coefficient found from formulas (18), (19) based on 90-day historical volatility data (b).

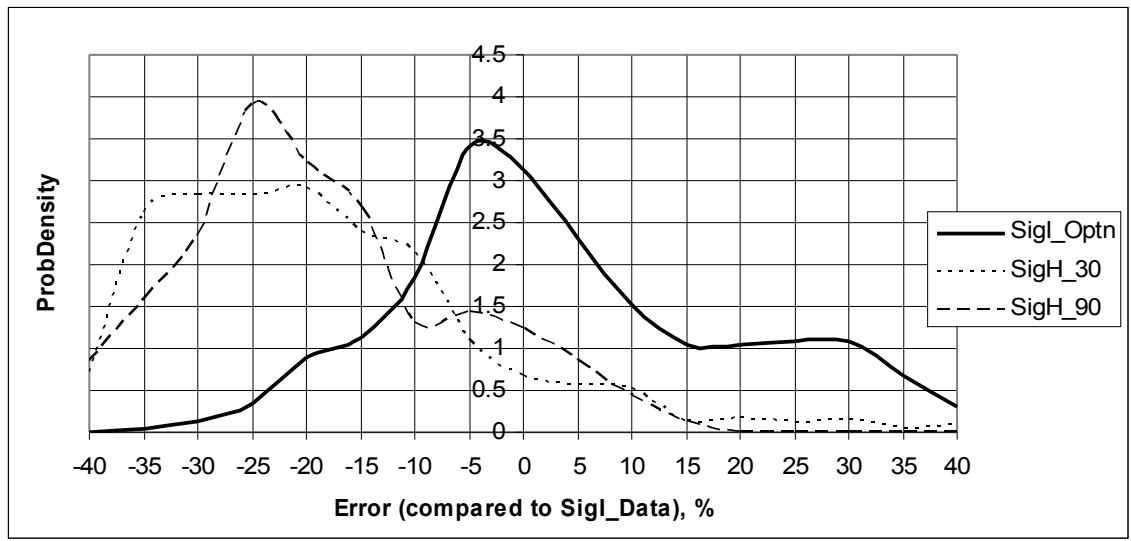

Fig 2. The error distribution for BSM and proposed "mirror-time" diffusion discount model calculated according to (20) compared to implied volatility VXO index *SigI_Data*. Average error for the diffusion-discount model results in terms of BS implied volatility *SigI_Model* is 1.85%; for BSM results based on 30-, 90-day historical volatility - 21.15% and 21.13%.

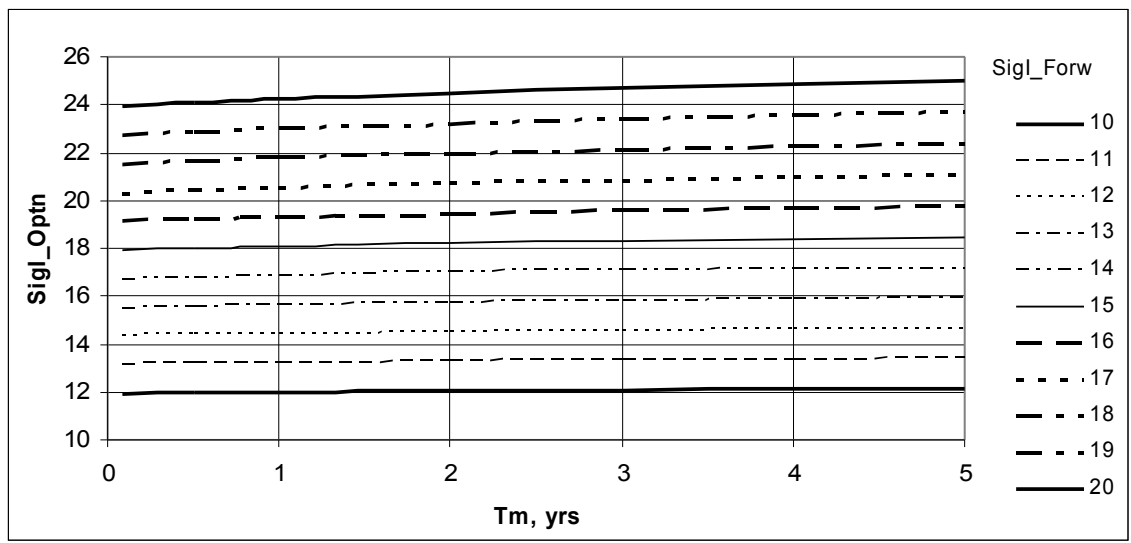

a

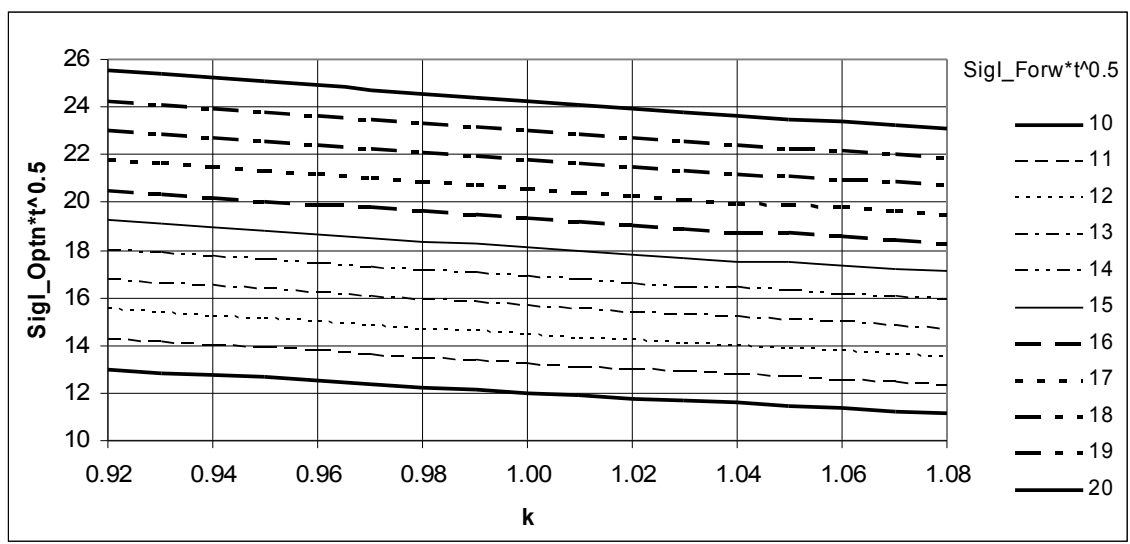

b

Fig.3. Proposed diffusion-discount model results calculated for caps according to (20) and expressed in terms of Black implied volatilities *SigI_Model* for different maturities and forward interest rate-based volatilities *SigI_Forw*: dependencies of maturity $T_m$ (a) and relative moneyness $k$ (b).